\documentclass[iop]{emulateapj}
\slugcomment{{\sc Accepted by ApJ:} June 11th, 2022.} 
\usepackage{amsmath}
\usepackage{mathtools}
\usepackage{hyperref}
\usepackage[switch]{lineno}
\usepackage{polski}
\renewcommand{\abstractname}{Abstract}
\usepackage{perpage}
\MakePerPage{footnote}
\def\apjs{{\rm ApJS}}                  
\def\apjs{{\rm ApJ}}                   
\def\apjl{{\rm ApJ Let}}               
\def\Msol{\thinspace\hbox{$\hbox{M}_{\odot}$}}
\def\Zsol{\thinspace\hbox{$\hbox{Z}_{\odot}$}}

\def\ie{{\it i.e.} }                    
\def\eg{{\it e.g.} }                    
\def\a4{\hsize 17.0cm \vsize 25.cm}

\defcitealias{Kroupa2001}{Kroupa}
\defcitealias{MartinezGonzalezetal2018}{I}
\defcitealias{MartinezGonzalezetal2019}{II}
\shorttitle{Dust in Low-Metallicity Molecular Clouds}
\shortauthors{Mart\'{i}nez-Gonz\'alez  et al.}

\begin{document}

\title{Dust Grain Growth \& Dusty Supernovae in Low-Metallicity Molecular Clouds}

\author{Sergio Mart\'{i}nez-Gonz\'{a}lez\altaffilmark{1},
   Richard W\" unsch\altaffilmark{2},
   Guillermo Tenorio-Tagle\altaffilmark{3},
   Sergiy Silich\altaffilmark{3},
   Dorottya Sz\'{e}csi\altaffilmark{4,5}, 
   Jan Palou\v s\altaffilmark{2}}

\altaffiltext{1}{CONACYT-Instituto Nacional de Astrof\'isica, \'Optica y Electr\'onica, AP 51, 72000 Puebla, M\'exico: sergiomtz@inaoep.mx} 
\altaffiltext{2}{Astronomical Institute, Czech Academy of Sciences,  Bo\v{c}n\'\i\ II 1401/1, 141 00 Praha 4, Czech Republic}
\altaffiltext{3}{Instituto Nacional de Astrof\'isica, \'Optica y Electr\'onica, AP 51, 72000 Puebla, M\'exico} 
\altaffiltext{4}{I. Physikalisches Institut, Universit\"at zu K\"oln, Z\"ulpicher Strasse 77, D-50937 K\"oln, Germany}
\altaffiltext{5}{Institute of Astronomy, Faculty of Physics, Astronomy and Informatics, Nicolaus Copernicus University, Grudzi\k{a}dzka 5, 87-100 Toru\'{n}, Poland}

\begin{abstract} 
We present 3-D hydrodynamical models of the evolution of superbubbles powered by stellar winds and supernovae from young coeval massive star clusters within low metallicity ($Z = 0.02\,\Zsol$), clumpy molecular clouds. We explore the initial stages of the superbubble evolution, including the occurrence of pair-instability and core-collapse supernovae. Our aim is to study the occurrence of dust grain growth within orbiting dusty clumps, and in the superbubble's swept-up supershell. We also aim to address the survival of dust grains produced by sequential supernovae. The model accounts for the star cluster gravitational potential and self-gravity of the parent cloud. It also considers radiative cooling (including that induced by dust) and a state-of-the-art population synthesis model for the coeval cluster. As shown before, a superbubble embedded into a clumpy medium becomes highly distorted, expanding mostly due to the hot gas streaming through low density channels. Our results indicate that in the case of massive ($\sim10^7\,\Msol$) molecular clouds, hosting a super star cluster ($\sim5.6\times10^5\,\Msol$), grain growth increments the dust mass at a rate $\sim4.8\times10^{-5}\,\Msol$ yr$^{-1}$ during the first $2.5$\,Myr of the superbubble's evolution, while the net contribution of pair-instability and core-collapse supernovae to the superbubble's dust budget is $\sim1200\,\Msol (M_{SC}/5.6\times10^{5}\,\Msol)$, where $M_{SC}$ is the stellar mass of the starburst. Therefore, dust grain growth and dust injection by supernovae lead to create, without invoking a top-heavy initial mass function, massive amounts of dust within low-metallicity star-forming molecular clouds, in accordance with the large dust mass present in galaxies soon after the onset of cosmic reionization.
\end{abstract}


\section{Introduction} 
\label{sec.intro}

Most massive stars reside within young massive star clusters \citep[]{LadaandLada2003,Portegies-Zwartetal2010}. There, hundreds to thousands of massive stars inject large amounts of metal-enriched gas via stellar winds and supernova explosions. Consequently, a localized dust enrichment of galaxies originating from massive star clusters can be expected \citet{Consiglioetal2016}. An important amount of this dust is likely produced after the efficient condensation of the ejecta of core-collapse supernovae \citep[e.g.][ and references therein]{TodiniandFerrara2001,Indebetouwetal2014} and pair-instability supernovae \citep[PISNe,][]{Nozawaetal2003,Cherchneff2010}. A competing mechanism that may massively enhance the amount of dust is the accretion of gas species onto already existing dust grains  \citep{Dwek1998,Zhukovskaetal2008,Asanoetal2013,Caluraetal2008,Caluraetal2014}. Both processes, dust produced by supernovae and dust grain growth, are expected to play an important role in local and high-redshift galaxies. For instance, \citet{GallandHjorth2018} concluded that the amount of dust observed in high-redshift galaxies comes from an efficient supernova dust production, followed by a rapid dust grain reformation if dust is destroyed by high-velocity shocks \citep[see also][]{Priestleyetal2021}. Observations of Ly$\alpha$ systems and quasar host galaxies at high redshifts suggest a rapid transition (on the order of a few Myr) from having a dust-poor to a dust-rich interstellar medium \citep[\eg ][]{Michalowskietal2010,Mattssonetal2015}, particularly if star formation took place with a top-heavy initial mass function \citep{GallandHorth2011,DwekandCherchneff2011}. It is also worth mentioning that the cold envelopes of the most massive AGB stars could also be important dust producers at high-redshifts \citep[\eg][]{Valianteetal2009,LesniewskaandMichalowski2019}. Moreover, dust formation from early-type carbon-rich Wolf--Rayet binaries could also represent an important source of dust in low-metallicity environments \citep[$Z\leq0.65\,\Zsol$][]{Lauetal2021}. 

The three-dimensional hydrodynamical evolution of dusty supernova remnants originating within young massive stellar clusters was explored in \citet{MartinezGonzalezetal2018}. For that purpose, \textsc{Cinder (Cooling INduced by Dust \& Erosion Rates)}, a module for the adaptive mesh refinement code \textsc{Flash} \citep{Fryxelletal2000}, was developed. 
With \textsc{Cinder} one can follow the survival rate of dust condensed out of a supernova ejecta and subjected to multiple shock processing, that includes the passage of the reverse shock and its bouncing back, the interaction with shocked stellar winds and the crossing of sequential supernova forward shocks. In the case of off-centered supernova remnants, facing a steep density gradient, a blowout phase is triggered: the supernova remnants elongate in preferential directions, become Rayleigh--Taylor unstable and the supernova ejecta suffers a rapid decline in density and temperature \citep{TenorioTagleetal2015b,Jimenezetal2021}. The main result of the ``pyroclastic blowout model'' is that clustered supernova explosions are likely to cause a net increase in the amount of dust in the free wind region surrounding their parental stellar clusters. 

However, the mechanical feedback provided by the cluster leads to the creation of a wind-blown superbubble \citep[see][]{Weaveretal1977, MacLowMcCray1988, KooMcKee1992,BKSilich1995}, where a supershell of swept-up interstellar matter encompasses the hot and tenuous free/shocked wind region. Thus, one cannot exclude that supernova remnants originating from the star cluster overrun the free and shocked wind regions, to then impact directly the swept-up supershell. This results in a further processing of both, the ejecta dust and the dust grains present in the supershell. Furthermore, supershells represent a fertile environment for the growth of dust grains even in high-redshift ($z\sim6$) galaxies at supersolar metallicities \citep{MartinezGonzalezetal2021}.

The collision of supernova remnants with a wind-driven shell has been previously explored \citep{TenorioTagleetal1990,TenorioTagleetal1991,Francoetal1991,Rozyczkaetal1993,Dwarkadas2005,Dwarkadas2007,vanMarleetal2015,Haidetal2016}. In \citet{MartinezGonzalezetal2019}, it has been shown that the pre-existent dust locked-up in a wind-driven shell is mostly not in danger of being significantly destroyed after the collision of blast waves with the surrounding shell. This is expected as blast waves
would transmit weakly into the much denser encompassing shell.

In the present work, we extend the models presented in \citep[][hereafter referred as Papers I and II]{MartinezGonzalezetal2018,MartinezGonzalezetal2019}, and follow the wind-blown superbubble evolution in the context of low-metallicity clumpy molecular clouds. Those low-metallicity environments may be representative of the Green Pea galaxies, a local analog class of high redshift Ly$\alpha$ emitting galaxies with a high star formation rate \citep{Cardamoneetal2009,Michevaetal2017,Svobodaetal2019,Franecketal2021}. They are also interesting since an increased pre-supernova feedback (harder ionizing radiation and increased photon fluxes) is expected in low-metallicity environments \citep[\eg][]{McLeodetal2021}. We will consider the potential well of the molecular cloud and the star cluster, the growth of grains originally residing within clumps, as well as dust grains produced in multiple pair-instability and core-collapse supernova explosions. 

The Paper is organized as follows: In Section \ref{sec:models} we describe the characteristics of the host molecular cloud (\ref{sec:clumpy}), the central star cluster and the wind-driven superbubble (\ref{sec:ssc}), the growth of dust grains (\ref{sec:growth}), and the occurrence of supernovae (\ref{sec:supernovae}). In Sections \ref{sec:results}-\ref{sec:PISN_results}, we present the results of the hydrodynamical simulations regarding the early evolution of the superbubbble, and the dust mass evolution.
In Section \ref{sec:limitations} we highlight the limitations of our model. Finally, in Section \ref{sec:concluding} we summarize our results and present the main conclusions.

\section{Model}
\label{sec:models} 

\subsection{Clumpy Molecular Cloud}
\label{sec:clumpy}

We have modelled a hierarchical molecular cloud consisting of a collection of small clumps and a tenuous interclump envelope \citep{TenorioTagleetal2006,DraineBook2011}. It is assumed that the cloud's metallicity is $Z=0.02\,\Zsol$, and that the cloud's clumpy density field is determined by \citep{Reynoldsetal2019}

\begin{equation}
  \label{eq:rho0}
  \rho_{cl}({\bf x}) = \rho_0\left(1 + 
    \sum_{i=1}^{n_{cl}} \frac{\rho_i}{\rho_0} e^{-2(\|{\bf x}-{\bf x}_i\|/r_i)^2}\right),
\end{equation}

where ${\bf x}=(x,y,z)$, $\rho_0$ is the interclump gas mass density, $n_{cl}$ is the number of clumps, $\rho_i$ and $r_i$ are the mass density and radius of the $i$th spherical clump centered at ${\bf x}_i=(x_i,y_i,z_i)$. Hereafter, we assume that $n_{cl} = 1000$ and adopt for the interclump gas density $\rho_0 = 2.128 \times 10^{-22}$\,g\,cm$^{-3}$, with
a mean mass per ion $1.22 m_{H}$, and a mean mass per molecular gas $2.33 m_{H}$, where $m_{H}$ is the proton mass.

Each clump is initially assumed to have a turbulent velocity field, as implemented by \citet{Tayloretal2018}. The clumps' kinetic energy spectra take the form $E(k)\propto k^{\alpha}$, with wavenumbers $k$ between $2\times2\pi/L$ and $32\times2\pi/L$, where $L$ is the length of the computational domain, and $\alpha=-5/3$. The virial ratio of the clumps is assumed to be $0.9$. We have taken normally-distributed values of $x_i$, $y_i$, $z_i$, $\rho_i/\rho_0$ and $r_i$. The mean and standard deviation of $x_i$, $y_i$, and $z_i$ are $0$\,pc and $20$\,pc. For $\rho_i/\rho_0$ they are $66$ and $50$, respectively. For $r_i$ these are $1.0$\,pc and $0.76$\,pc, respectively. The clumps amount to $M_{cl}\sim3.73\times10^6\,\Msol$ of molecular gas (roughly one third of the total gas mass $\sim1.03\times10^7\,\Msol$), with $\sim68$ per cent of the clumps located within $13$\,pc. 

\subsection{The Star Cluster \& the Wind-driven Superbubble}
\label{sec:ssc}

We have set-up the three-dimensional hydrodynamical evolution of a superbubble within a molecular cloud with the characteristics described in Section \ref{sec:clumpy} (see Table \ref{tab:1}), using a similar setup to that described in Papers \citetalias{MartinezGonzalezetal2018} and \citetalias{MartinezGonzalezetal2019}. The superbubble is driven by the mechanical feedback of a coeval young massive stellar cluster located at the cloud's center. To this end, we have used the adaptive mesh refinement code FLASH v4.3 \citep{Fryxelletal2000}. The employed hydrodynamical solver is a modified version of the Piecewise Parabolic Method introduced by \citet[][]{ColellaandWoodward1984}. Similarly to Paper \citetalias{MartinezGonzalezetal2018}, the simulations take into account the star cluster's gravitational potential and the self-gravity of the gas calculated by the tree-based solver developed by \citet{Wunschetal2018}, and the optically thin cooling function for gas at temperatures $T\geq 10^4$\,K \citep{Schureetal2009}, and for gas at temperatures $T < 10^4$\,K, gas cooling  due to 
radiation emitted from hydrogen-deuteride ground-state rotational transition \citet{DalgarnoandMcCray1972}. We have used \textsc{Cinder}\citep{MartinezGonzalezetal2018,MartinezGonzalezetal2019} to follow the injection and destruction of dust grains due to thermal sputtering, as well as the cooling induced by gas-grain collisions in hot plasmas ($T \gtrsim 10^5$\,K). As in Papers \citetalias{MartinezGonzalezetal2018} and \citetalias{MartinezGonzalezetal2019}, we have not included other grain destruction mechanisms such as kinetic sputtering and grain shattering (see subsection \ref{sec:limitations}). Additionally, the process of dust grain growth is incorporated into \textsc{Cinder} for the first time. 

\begin{deluxetable*}{lccccccccccc}
\tablecolumns{9} \tablewidth{0pc}
\tablecaption{\label{tab:1} \sc The Host Molecular Cloud \& the Star Cluster}
\tablehead{
\colhead{Model}& \colhead{$M_{MC}$}& \colhead{$Z$} & \colhead{$n_{cl}$} & \colhead{$\rho_{0}$} & \colhead{$M_{SC}$}&                                   \colhead{$R_{ch}$} & \colhead{$R_{SC}$} & \colhead{$\epsilon$} & \colhead{No. Stars} & \colhead{Grain Growth} & \colhead{Ejecta dust} \\
&\colhead{\tiny $\Msol$}& \colhead{\tiny \,$\Zsol$} & \colhead{\tiny $-$} &  \colhead{\tiny\,g\,cm$^{-3}$} & \colhead{\tiny $\Msol$}& \colhead{\tiny\,pc} & \colhead{\tiny\,pc} & \colhead{\tiny per cent} & \colhead{\tiny $\geq 71\,\Msol$} & $-$ & $-$ \\
}
\startdata
\textsc{GrainGrowth}   &$10^{7}$ & $0.02$ & $1000$ & $2.26\times10^{-22}$ & $5.6\times10^{5}$ & $1.0$ & $3.0$  & $3.7$ & $186$ & $\mbox{Yes}$
& $\mbox{No}$ \\
\textsc{SNDust}  &$10^{7}$ & $0.02$ & $1000$ & $2.26\times10^{-22}$ & $5.6\times10^{5}$ & $1.0$ & $3.0$  & $3.7$ & $186$ & $\mbox{No}$
& $\mbox{Yes}$
\enddata 
\tablecomments{The Table summarizes the main properties of the host molecular cloud (mass, metallicity, number of molecular clumps and the interclump mass density), and the star cluster (stellar mass, the stellar distribution's characteristic scale and cut-off radii, the global star formation efficiency, and number of progenitor stars that end as PISNe). The last two columns indicate whether
the processes of dust grain growth and dust injection by supernovae are considered.}
\end{deluxetable*}

\begin{figure*}
\epsscale{1.2}
\plotone{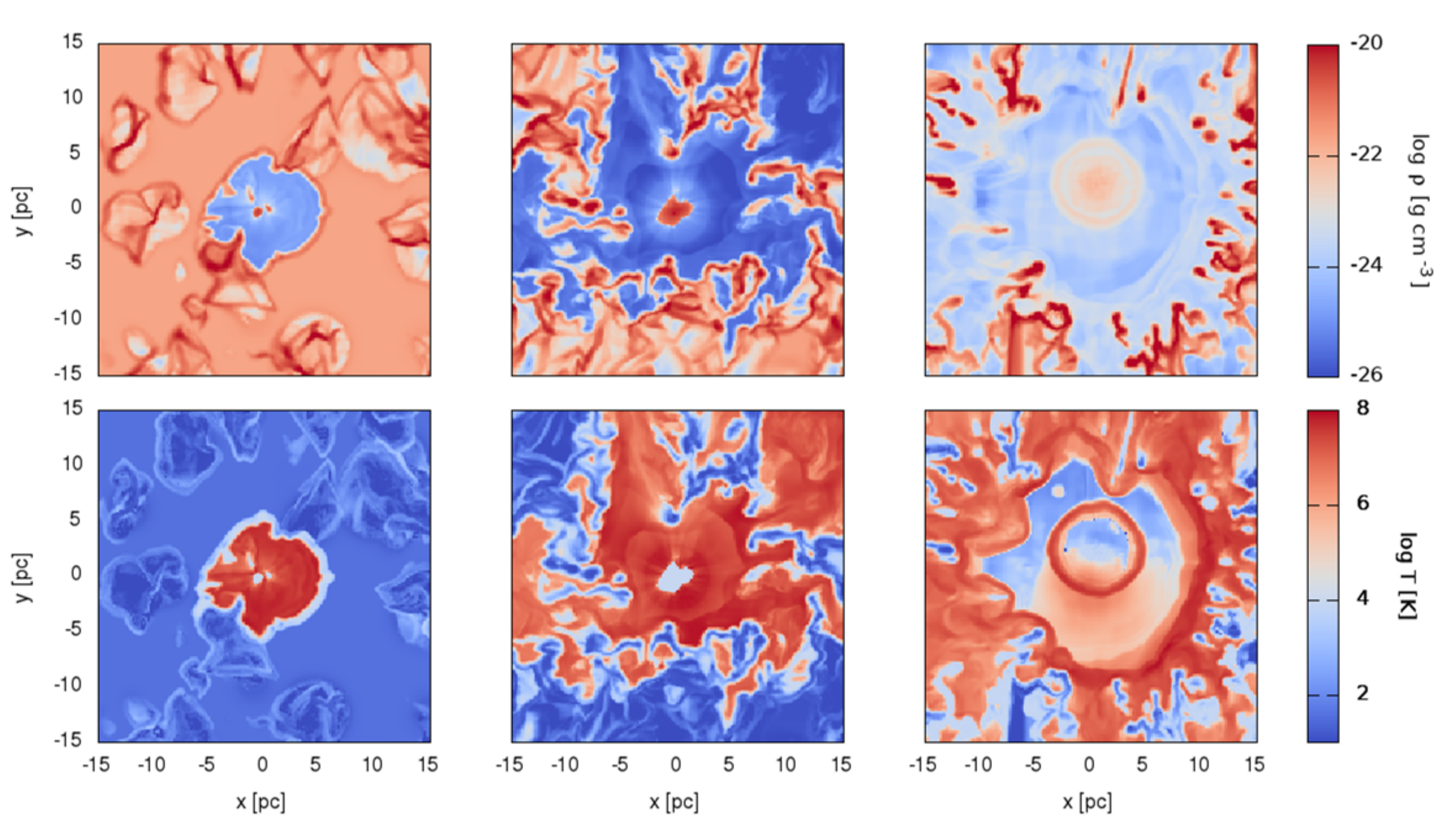}
\caption{The stream of hot gas through low-density channels. The upper and bottom panels display slides in the $z$-plane of the distribution of gas mass density and gas temperature, respectively, at three different times (left: $100$\,kyr; center: $2.0$\,Myr; right: $2.65$\,Myr). The flow of thermalized matter between low-temperature overdensities leads first to the distortion of the starburst-driven shell, and then to the development of multiple ragged shells at later times. Note the left and central panels correspond to model \textsc{GrainGrowth}, while the right panels depict model 
\textsc{SNDust}, once PISNe are occurring and the superbubble's interior is filled with wind and supernova matter. At $2.65$ Myr, a young supernova remnant can be identified as a ring-like hot structure exhibiting its forward and reverse shocks, and surrounded by the free-wind region with a density that falls as $\sim r^{-2}$. Further out, there is a global reverse shock thermalizing the free cluster wind and causing a temperature larger than $10^7$ K (appearing as a red zone at $\sim 10$ pc from the center). Much further out there is a leading shock that has managed to display the 
original fragments from the central regions, while advancing with different speeds in different directions given the uneven density left in the surrounding medium.}
\label{Figure:1}
\end{figure*}

The simulations are inscribed into cubes ($40$\,pc)$^3$ and ($140$\,pc)$^3$ (when indicated) in a grid $256^3$ in Cartesian geometry, with outer boundary conditions set to outflow. 

We follow the evolution of the star cluster wind mechanical output, mass-loss rate and terminal speed using the \textsc{BoOST} stellar model grids and the \textsc{SynStars} stellar population synthesis code \citep{Szecsietal2020,Franecketal2021}. In particular, we rely on a stellar grid with  metallicity $0.02\,\Zsol$ (originally presented by \citealt{Szecsietal2015}), which consists of slowly rotating single stars computed with the `Bonn' stellar evolution code with standard wind mass-loss prescriptions (\citet{Vinketal2000,Vinketal2001} type mass loss in the OB phase and \citet{NieuwenhuijzenanddeJager1990} mass loss in the supergiant phase). As discussed by \citet{SzecsiandWunsch2019}, these models spend their main-sequence lifetimes as hot OB stars, and their post-main-sequence as cool supergiants; none of them form Wolf--Rayet stars, as the metallicity is too low for the self-stripping by the wind. The stellar winds of these massive stars do contribute to the mechanical energy inserted into the cluster, but neither the wind yields, nor the supernova yields from these models are included into the 3--D hydrodynamical simulations. 

We apply a \citetalias{Kroupa2001} initial stellar mass function (IMF) in the mass interval $[0.01$\textendash$120]\,\Msol$. The adopted stellar mass\footnote[1]{This value is inspired by the derived mass of NGC\,604 \citep[][]{Relanoetal2016}. As we shall see, the simulated star cluster, superbubble and molecular cloud resemble the complex morphology of massive H\textsc{II} regions such as NGC\,604.} is $5.6\times 10^5\,\Msol$, distributed according to a Schuster stellar density profile \mbox{$\rho_\star \propto [1+(r/R_{ch})^2]^{-\beta}$} with characteristic scale radius $R_{ch}=1$\,pc, cut-off radius $R_{SC}=3$\,pc and index $\beta$=1.5 \citep[][]{Palousetal2013,MartinezGonzalezetal2016,MartinezGonzalezetal2017}. The global star formation efficiency is $\epsilon=M_{SC}/M_{MC}=3.7$ per cent.

\subsection{Dust Grain Growth}
\label{sec:growth}

\begin{figure}
\epsscale{1.2}
\plotone{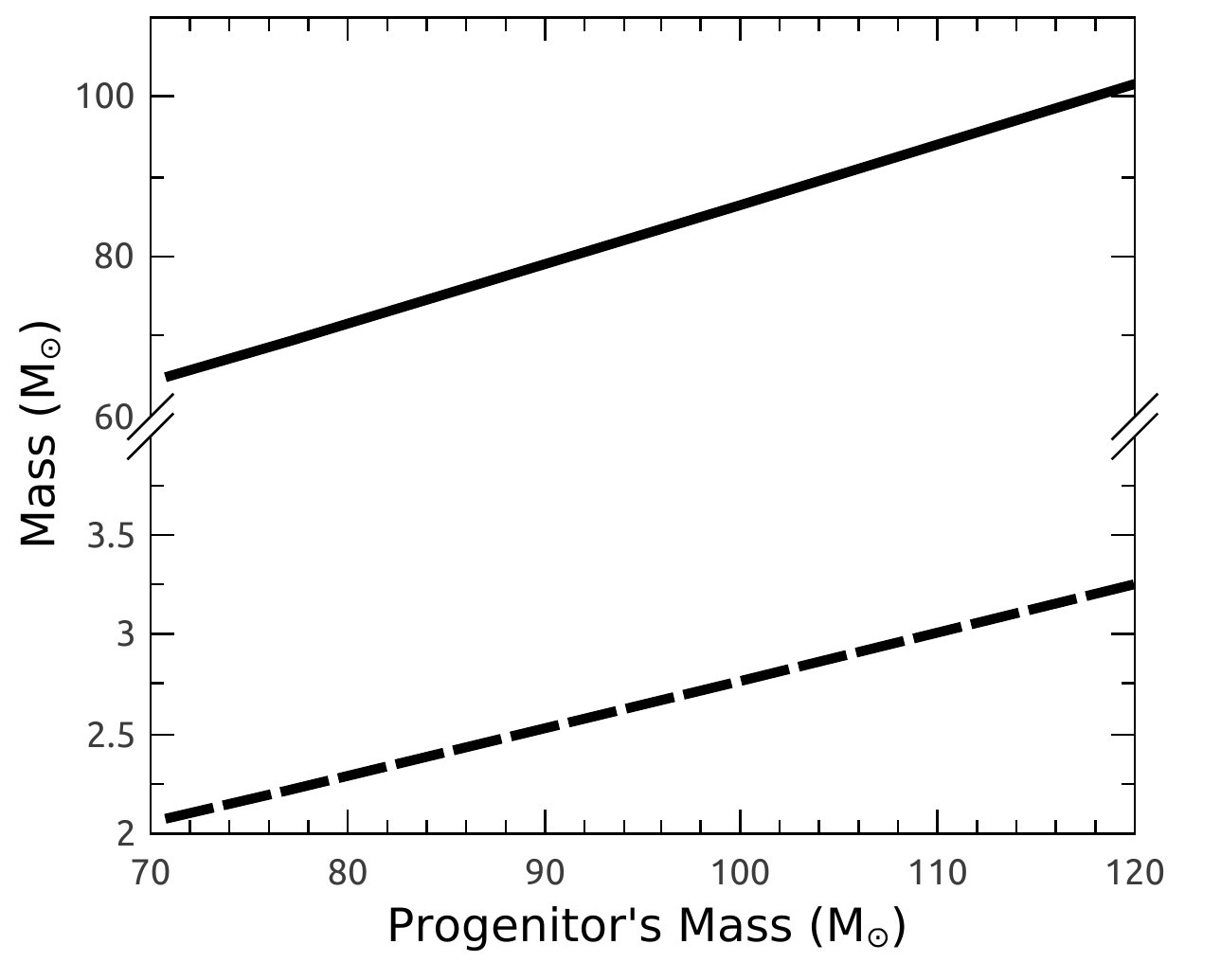}
\caption{The plot shows the ejecta mass (solid line) and the assumed ejecta-condensed dust mass (dashed line) as functions of the progenitor's mass (measured at the zero-age main-sequence). The mass of the ejecta corresponds to the progenitor's final mass, and the ejecta-condensed dust mass is assumed to be $3.2$ per cent of the ejecta mass \citep{Cherchneff2010}.}
\label{Figure:2}
\end{figure}

\begin{figure*}
\epsscale{1.2}
\plotone{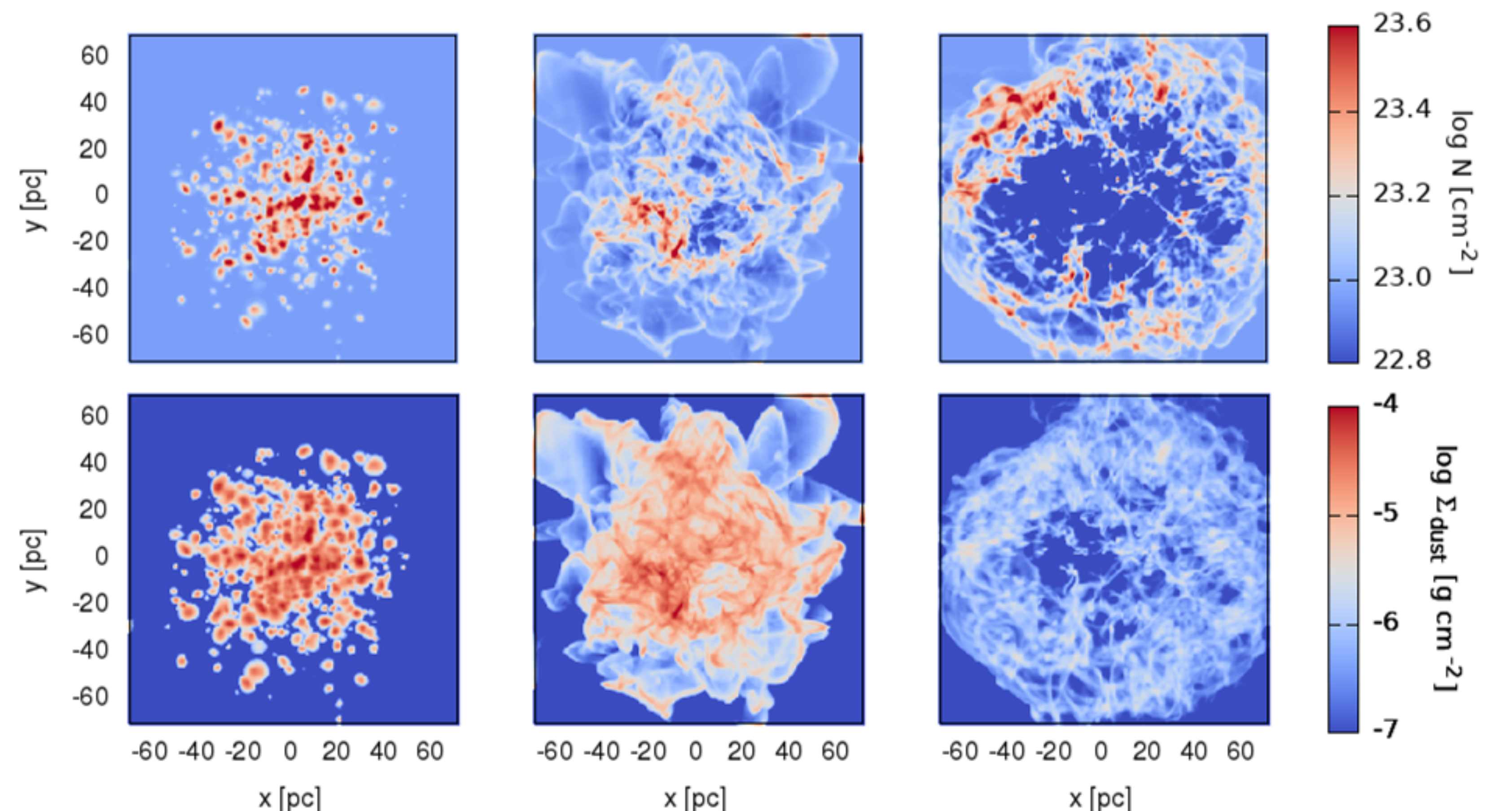}
\caption{The wind-blown superbubble's evolution in an initially clumpy medium. The central star cluster has a mass $5.6\times 10^5\,\Msol$. The upper, and bottom panels show the gas column density, and the dust mass surface density (model \textsc{GrainGrowth}), respectively. The left and central and right panels correspond to three different evolutionary times (left: $100$\,kyr, center: $2.5$\,Myr, for model \textsc{GrainGrowth}; right: $3.2$\,Myr for model \textsc{SNDust}). Soon, the host molecular cloud and the supperbubble acquire complex morphologies due to the action of gravity and the flow of hot gas through channels between clumps/filaments. An animated version of this Figure can be found on \href{https://user-images.githubusercontent.com/90579577/170849006-426f50e6-2135-4c85-82e3-c24747a46bec.mp4}{Github}.}
\label{figure:3}
\end{figure*}

We have incorporated the process of dust grain growth via the accretion of gas species \citep{Spitzer1978}, so individual dust grains increase their mass at a rate $\dot{m}_{gr} = 3 m_{gr}\dot{a}/a$, where $m_{gr}$ is the grain mass and $\dot{a}$ is the rate of increase in their radius $a$. Similarly to Paper \citetalias{MartinezGonzalezetal2019}, 
we consider graphite and silicate dust grains and a log-normal grain size distribution of the form $\sim a^{-1} \exp \{-0.5 [\log(a/a_{0})/\sigma]^2\}$; with $a_{0}=0.1$\,$\mu$m and $\sigma=0.7$ and minimum and maximum grain sizes $a_{min}=5$\,nm and $a_{max}=0.5$\,$\mu$m, respectively. The grain size distribution is sampled by $10$, logarithmically-spaced, size bins. The dust-to-gas mass ratio, $\mathcal{D}^{(i)}$, in a bin with representative grain size, $a_{m}^{(i)}$, at time $t + \Delta t$ is \citep{ValentiniandBrighenti2015}

\begin{eqnarray}
 \mathcal{D}^{(i)}(t+\Delta t) &=& \mathcal{D}^{(i)}(t) 
  \left( 1 + \displaystyle \frac{3|\dot{a}|\Delta t}{a_{m}^{(i)}} \right) ,
\end{eqnarray}

where $\dot{a}$ is given by

\begin{eqnarray}
\label{eq:dota}
 \dot{a} = \frac{f_{gr}\rho_{r} S}{4 \rho_{gr}} \left[f_{M_{d}}^{(i)}\mathcal{D}^{max} -\mathcal{D}^{(i)}\right]  \left(\frac{8 k_B T}{\pi  \mu_{r}}\right)^{1/2} .
\end{eqnarray}

In equation \eqref{eq:dota} $\rho_{r}$ is the local refractory gas mass density, $\rho_{gr}$ is the grain's bulk density, $S$ is the sticking coefficient, $f_{gr}$ is the fraction of grain species (for simplicity $f_{gr}$ is set to $0.5$ for both graphite and silicate grains), $f_{M_{d}}^{(i)}$ is the initial dust mass fraction of the $i$th grain size, $k_B$ and $\mu_{r}=18.11m_{H}$ are the Boltzmann constant and the mean mass per refractory atom (\ie excluding non-condensable gas species). $\mathcal{D}^{max}=2.1\times10^{-4}$ is the dust-to-gas mass ratio in the case all refractory elements are locked-up onto dust grains. This value is taken from the initial mass fractions of refractory elements in the `Bonn' stellar evolution code. At the explored molecular gas metallicity ($Z = 0.02\,\Zsol$), refractory elements make up a gas mass fraction of $\lesssim10^{-6}$, a quantity that is well approximated by the relation  $\mathcal{D}^{max} \approx 10^{-2}\times Z$ \citep{MartinezGonzalezetal2021}. 

We allow dust grain growth to occur at gas temperatures $T\leq 1000$\,K \citep{Nozawaetal2012}. At these temperatures, grains with radius $\geq 5$ nm embedded into a gas with molecular gas mass density $\leq 10^5$\,cm$^{-3}$ have an equilibrium temperature $\leq 30$\,K. Consequently, the grain sticking coefficient $S$ is approximated as $1.0$ \citep{Ferraraetal2016}. 

\subsection{Supernovae}
\label{sec:supernovae}

At their oxygen-burning phase, massive CO stellar cores ($\gtrsim 65\,\Msol$) undergo the so-called electron-positron pair-creation instability \citep{Fowleretal1964}. This leads to the disruption of the whole star as a PISN, leaving behind neither a black hole, nor a neutron star \citep[][and references therein]{Kozyrevaetal2014}. According to the \textsc{BoOST} low-metaillicity stellar tracks, that occurs for stars with masses $\gtrsim 71\,\Msol$ at the zero-age main-sequence.
 From the assumed IMF, we thus follow the occurrence of $186$ PISNe between $2.55$ and $3.18$\,Myr. The supernova rate is a function of the star cluster mass, the IMF, and the metallicity of the parent cloud, and thus evolves with time. We, however, have set a constant supernova rate for progenitors in the mass intervals $[71$\textendash$80)\,\Msol$, $[80$\textendash$90)\,\Msol$, $[90$\textendash$100)\,\Msol$, $[100$\textendash$110)\,\Msol$, and $[110$\textendash$120]\,\Msol$, with one supernova per $4250$ years, $3300$ years, $3050$ years, $2880$ years, and $2460$ years, respectively. Supernovae are randomly-distributed following the assumed stellar density profile. As PISNe completely obliterate their progenitor stars, their ejecta masses, $M_{ej}$, correspond to the final masses obtained from the \textsc{BoOST} stellar model grids. With respect to the total energy released in a single PISN event, we assume $E_{SN}=5\times10^{51}$\,erg. This energy is inserted into a sphere of radius equivalent to five grid-cells. Massive stars with final masses $\sim[10$\textendash$65]\,\Msol$ explode as core-collapse supernovae \citep{Szecsietal2020}. We have thus explored the occurrence of the first $50$ core-collapse supernovae starting after the last PISNe. Our assumption is that each core-collapse supernova releases $M_{ej} = 10\,\Msol$ of gas and $E_{SN}=10^{51}\,\mbox{erg}$ in the form of kinetic energy. 

For each supernova, the initial ejecta gas mass density distribution is assumed to be given by

\begin{eqnarray}
\label{eq:ejecta}
 \rho_{ej}=\frac{(3-\omega)}{4\pi}\frac{M_{ej}}{R_{SN}^3}\left(\frac{R_{SN}}{r'} \right)^\omega  ,
\end{eqnarray}

where $r'$ is the radial distance from the center of the explosion and $\omega=5/2$ was set for all supernovae. The initial ejecta velocity distribution is assumed to be

\begin{eqnarray}
\label{eq:vel}
 v_{ej} = \left( 2\frac{(5-\omega)}{(3-\omega)}\frac{E_{SN}}{M_{ej}} \right)^{1/2}\left(\frac{r'}{R_{SN}} \right) .
\end{eqnarray}

\section{Large-Scale Evolution}
\label{sec:results}

In our first model \textsc{GrainGrowth}, we
follow the occurrence of dust grain growth during the evolution of the molecular cloud and the superbubble driven by the central star cluster. 
Given the low gas temperatures involved in the host molecular cloud, and hence short cooling timescales and small cooling lengths, we have modelled the central ($40$\,pc)$^3$ at three spatial resolutions, $0.15$ pc and $0.31$ pc, respectively. Additionally, we tested the results with a larger computational domain, ($140$\,pc)$^3$ at a resolution $0.54$ pc, which encompasses all the molecular clumps. In this model we initially assume that half the mass of refractory elements within molecular clumps are already depleted onto dust grains, so that their dust-to-gas mass ratio is $\mathcal{D}=\displaystyle\frac{1}{2}\mathcal{D}^{max}$ (and zero elsewhere). Thus the initial dust mass within molecular clumps is $\displaystyle \frac{1}{2}\mathcal{D}^{max}M_{cl} \approx 373\,\Msol$. The initial population of dust grains is assumed to have been originated in prior (very early) generations of supernovae \citep{Nozawaetal2012}. We stop our calculations just before the occurrence of the first PISNe. 

\begin{figure}
\epsscale{1.2}
\plotone{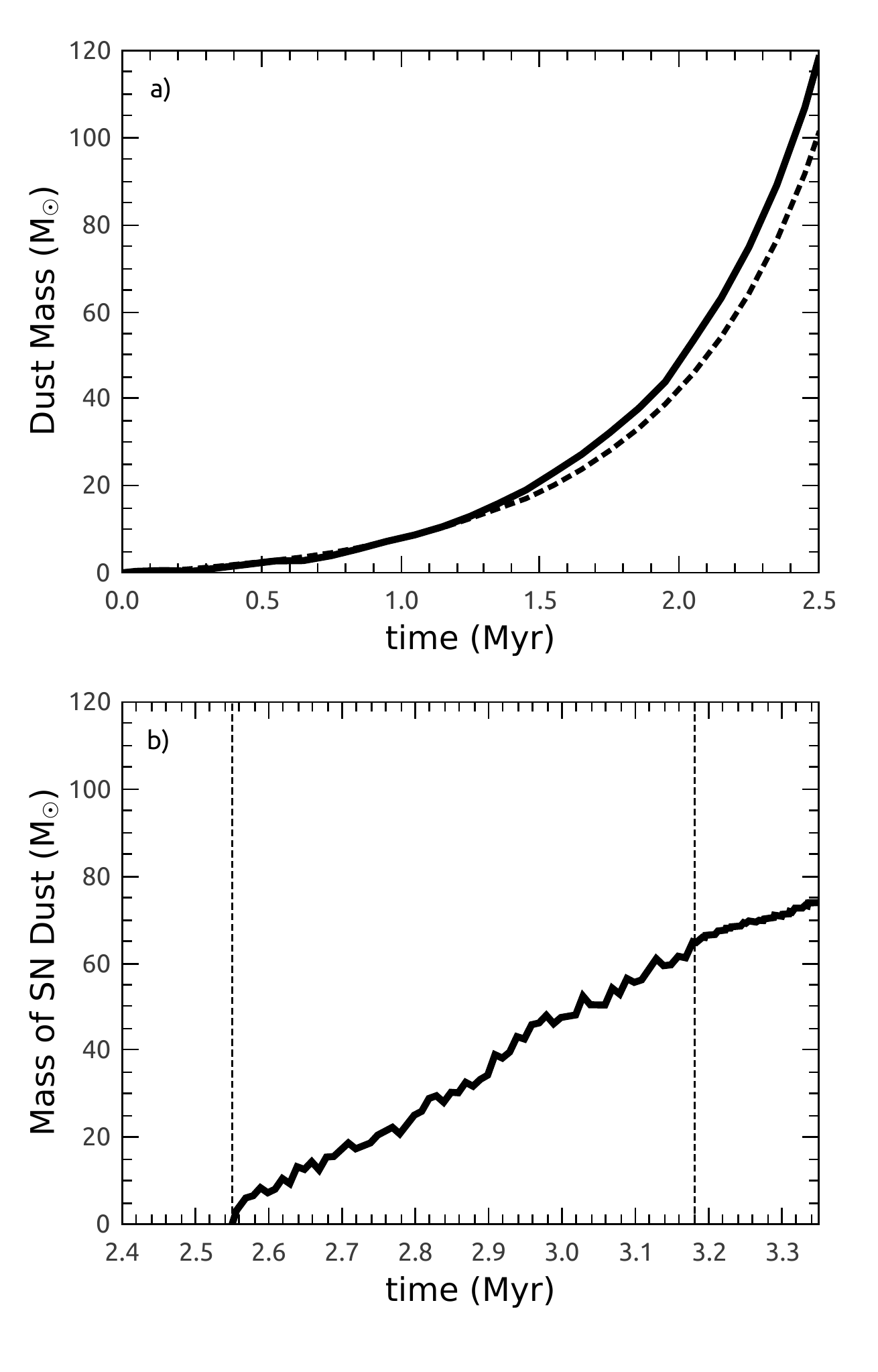}
\caption{The dust mass evolution. Panel {\it a} presents the increment of dust mass due to grain growth in the central part ($40^3$ pc$^3$) of the molecular cloud/supershell during the pre-supernova era (model \textsc{GrainGrowth}). The thick line depicts the result of the run with the $0.15$ pc spatial resolution, while the dashed line to the $0.31$ pc spatial resolution run. Panel {\it b} displays the cumulative ejecta-condensed dust mass deposited by PISNe. The vertical dashed lines in panel {\it b} mark the onset and finale of the PISN era. Note the different time intervals depicted in both panels.}
\label{figure:4}
\end{figure}

As the molecular cloud evolution proceeds, clumps become elongated and end up forming filamentary structures while they revolve around the gravitational potential well. We note that the potential well is dominated by the molecular gas, rather than the central star cluster. During the process, the clumps interact, collide and coalesce with other clumps \citep{Elmegreen1988}. The superbubble thus expands preferentially along the paths of least resistance, \ie through the lower density channels between clumps/filaments \citep[][see Figure \ref{Figure:1}]{TenorioTagleetal2006,Aluzasetal2012,RogersandPittard2013,Lucasetal2020,Lancasteretal2021a,Lancasteretal2021b}. 

The end result is that the superbubble morphology becomes highly distorted. The flow of thermalized matter between clumps/filaments leads first to the distortion of the star cluster wind-driven shell (displayed in the bottom panels
of Figure \ref{Figure:1} by the blue/cool contours encompassing the red/hot zones) and then to the development of multiple shells, some of them appearing as nested shells when they are seen projected against the background \citep[see Figures 2 and 3 in][]{TenorioTagleetal2006}. Those multiple shells are all interconnected, thus forming a large-scale supershell. 
Eventually, the superbubble, powered mostly by the SN shocks, will breakout/blowout from the host molecular cloud \citep[][]{TenorioTagle2002}: however this moment is not captured in our simulations within our computational domain.

The top and bottom panels in Figure \ref{figure:3} present the evolution of the gas column density, $N$, and dust mass surface density, $\Sigma_{dust}$, integrated along the line-of-sight at $100$\,kyr and $2.5$\,Myr  for model \textsc{GrainGrowth}. The right panels in the Figure correspond to $3.2$\,Myr in the case of model \textsc{SNDust}, so $\Sigma_{dust}$ corresponds to supernova-condensed dust only.

\section{Dust grain growth at low metallicity}
\label{sec:growth_results}

In model \textsc{GrainGrowth}, dust grain growth is promoted when the molecular clumps become filamentary and mix with the interclump material. It is also favoured when interclump gas is incorporated into the dense shells formed as the hot gas streams through low-density channels. We note that as the shells cool down quickly, a negligible amount of swept-up dust is destroyed. As a result, the dust mass increases at a rate $\sim4.8\times10^{-5}\,\Msol$ yr$^{-1}$, leading to a net increase of $\sim120\,\Msol$ within $2.5$\,Myr (see panel {\it a} in Figure \ref{figure:4}). The results with resolutions $0.31$ pc and $0.54$ pc show a reduction of $13$ and $25$ per cent, respectively, on the amount of dust grain growth. This is expected since decreasing the resolution  inhibits efficient mixing of dust-free and dust-rich gas via numerical viscosity.

As we have set the outer boundary conditions to outflow, there is some fraction of the dust mass that leaves the computational domains. However, the majority of the gas/dust overdensities remain within the central part of the molecular cloud.

We note that if dust-induced cooling of molecular gas were at play (not included in the current version of \textsc{CINDER}), the shells would be much thinner, denser and much colder, and thus they may promote a much more efficient growth of dust grains at higher gas metallicities \citep{MartinezGonzalezetal2021}. Higher resolution simulations are required in such a case.

\section{Dust injected by Supernovae}
\label{sec:PISN_results}

Model \textsc{SNDust} explores the injection of dust by sequential PISN and core-collpase supernovae, while molecular clumps are set to be dust-free (\ie no grain growth occurs). 
This model is studied at a resolution $0.54$ pc since the cooling length\footnote[2]{The cooling length can be defined as the product of the local sound speed and the cooling time, and increases with decreasing gas density and metallicity
\citep{Smithetal2017}.} in the cavity is very large due to its low density and high temperature. 

It is assumed that each PISNe leads to the condensation of $3.2$ per cent of the progenitor's final mass (around $3.25\,\Msol$ for a $120\,\Msol$ progenitor star) onto dust grains (see Figure \ref{Figure:2}). Such high value is consistent with theoretical expectations for dust condensation in PISNe occurring from zero-metallicity progenitors \citep{Cherchneff2010}. The dust mass condensed out of supernova ejecta for core-collapse supernovae is assumed as normally-distributed, with a mean of $0.6\,\Msol$ and standard deviation of $0.1\,\Msol$. Similarly to the grains originally locked-up within molecular clumps, the simulated supernovae are also assumed to inject graphites and silicates in equal mass fractions following a log-normal grain size distribution. We have not included the process of ion trapping onto dust grains \citep[see][]{Kirchschlageretal2020}, although such a process can produce a further growth of ejecta-condensed dust grains.

Panel {\it b} in Figure \ref{figure:4} shows that, despite the occurrence of several PISN explosions, the amount of ejecta-condensed dust increments steadily, at a rate $10^{-4}\,\Msol$ yr$^{-1}$ between $2.55$ and $3.18$\,Myr. To put it into perspective, we find that from $ 470\,\Msol$ of ejecta dust that were injected during the era of PISNe, about $13$ per cent ($\sim 61\,\Msol$) remains in the superbubble's cavity $\sim320$\,kyr after the last PISN. This value is in agreement with the conclusions raised in Paper \citetalias{MartinezGonzalezetal2018}, that explored, at a higher resolution ($\sim 0.11$ pc), the case of sequential core-collapse supernova explosions, occurring at a similar rate, within a star cluster wind. It also builds on the conclusion that massive shells represent insurmountable barriers that protect the vast majority of the pre-existing surrounding dust from being shock-processed (Paper \citetalias{MartinezGonzalezetal2019}). 

Progenitors with final masses $\sim[10$\textendash$65]\,\Msol$ explode as core-collapse supernovae. We have found that from the $\sim30\,\Msol$ of ejecta dust that were injected into the simulation, $\sim32$ per cent ($\sim9.5\,\Msol$) survives after $175,000$ years from the first core-collapse supernova explosion. 

If we extrapolate the rate of net dust injection by core-collapse supernovae until $\sim 21$ Myr of the cluster's evolution (the time needed to explode a $10\,\Msol$ star in the \textsc{BoOST} low-metaillicity stellar tracks), we can expect a net increase in the dust mass by $\sim1200\,\Msol (M_{SC}/5.6\times10^{5}\,\Msol)$. Applying this estimate to the gravitationally-lensed galaxy A2744\_YD4 \citep[at a redshift $z\sim8.38$,][]{Laporteetal2017}, with a total stellar mass $\sim2\times10^9\,\Msol$, we obtain a net dust contribution by supernovae $\sim4.3\times10^6\,\Msol$. This value fits pretty well the dust mass  ($6\times10^6\,\Msol$) derived by \citet{Laporteetal2017} (see also \citealt{Behrensetal2018}). Note that our estimate relies on a standard \citetalias{Kroupa2001} initial stellar mass function, \ie a top-heavy IMF was not necessary in order to explain such a large dust mass produced by pair-instability and core-collapse supernovae. We have also tested the occurrence of the first five PISNe at a resolution $0.15$ pc in a domain ($40$ pc)$^3$ (see the right panels in Figure \ref{Figure:1}), and the results agree within $15$ per cent.

\section{Radiative Heating, Interstellar Magnetic Field, \& Gas-Dust Coupling}
\label{sec:limitations}

Radiative heating provided by the central stellar cluster is not included into the simulations. We have, however, set a standard heating rate $2\times10^{-26}$ erg s$^{-1}$
\citep{KoyamaandInutsuka2002}. This does not prevent molecular gas from cooling to low temperatures provided a sufficient density. Additionally, the gas can also cool adiabatically, which occurs in low density regions behind shocks.

Our model does not incorporate interstellar magnetic fields which may play an important role in molecular cloud dynamics. Nevertheless, recent measurements of the magnetic field strength in molecular clouds show that gravitational forces may exceed those due to magnetic pressure gradients \citep{Crutcher2012,Crutcheretal2019}. The presence of an interstellar magnetic field could initially increase the supershell's thickness, thus increasing the timescale for dust grain growth within the supershell \citep{Ferriereetal1991}.

As already stated, we have not modelled other grain destructive processes that require a shock to kinematically-decouple the gas particles and dust grains, such as kinetic sputtering and grain shattering. However, the superbubble's interior is very tenuous and thus gas-grain and grain-grain encounters are not very frequent even in the case of large gas-grain and grain-grain relative velocities
\citepalias[see Appendix A.2 in Paper][]{MartinezGonzalezetal2019}. This fact, and the fact that sequential supernova shocks soon decay into sound waves within the thermalized superbubble's interior \citep{TenorioTagleandBodenheimer1988}, are likely to decrease the relative importance of such processes. In addition, betatron acceleration, which occurs when charged dust grains gyrate along magnetic field lines \citep{Shull1977}, is not likely to play a significant role within the superbubble's interior given that at high temperatures ($\gtrsim 2\times 10^5 $ K), dust grains tend to be neutral \citep{McKeeetal1987}.

\section{Concluding Remarks} 
\label{sec:concluding}

By conducting 3--D hydrodynamical simulations, we have explored the early ($\sim 3.2$\,Myr) evolution of a starburst-driven superbubble and the onset of PISNe in a low-metallicity ($Z=0.02\,\Zsol$), clumpy molecular cloud. Our main purposes were to study the process of dust grain growth at low-metallicities and the fate of dust grains condensed from the ejecta of pair-instability and core-collapse supernovae. The star cluster's mechanical feedback has been modeled using the state-of-the-art \textsc{BoOST} stellar model grids \citep{Szecsietal2020} and the \textsc{SynStars} stellar population synthesis code \citep{Franecketal2021}.

Dense clumps, that randomly move in the gravitational potential well, coalesce and mix with the interclump matter. At the same time, the hot gas that fills the superbubble's cavity flows through the channels in between overdensities, to then create a net of interconnected shells.
It was found that the mixing of clump-interclump material acts as a trigger for an important enhancement (at a rate $\sim4.8\times10^{-5}$ $\Msol$ yr$^{-1}$ during the first $2.5$\,Myr of the superbubble evolution) of the dust mass in the filamentary/clumpy molecular cloud and the supershell.

Around $\sim 2.5$\,Myr of the superbubble evolution, the most massive stars in the central star cluster start to explode as PISNe. Their forward shocks move into the shock-heated, low-density cavity, and soon they decay to sound waves \citep{TenorioTagleandBodenheimer1988}. This behavior might inhibit the destruction of the ejecta-condensed dust grains via non-thermal, shock-induced dust grain disruptive processes. Overall, $\sim 13$ and $\sim 32$ per cent of the dust masses injected by PISNe and core-collapse supernovae, respectively, survive even after being processed in multiple supernova collisions, with net dust injection rates of the order $10^{-4}\,\Msol$ yr$^{-1}$ and $5.4\times 10^{-5}\,\Msol$ yr$^{-1}$, respectively. The destruction of dust grains locked-up in the shell is also largely inhibited as the multiple supernova blast waves are marginally transmitted into the supershell; a result originally presented in Paper \citetalias{MartinezGonzalezetal2019}.

The net dust contribution by supernovae alone is sufficient to explain the large dust mass present in the gravitationally-lensed galaxy A2744\_YD4 (at a redshift $z\sim8.38$) reported by \citet{Laporteetal2017}, without the need to invoke a top-heavy IMF.

We thus have demonstrated that the processes of dust grain growth and dust injection by supernovae are both efficient pathways that lead to massive amounts of dust to be present in low-metallicity environments populated by young stellar clusters.

\section{Acknowledgements}

This study was supported by CONACYT-M\'exico research grant A1-S-28458. S.M.-G. also acknowledges support by CONACYT through project n.482 of the ``Programa Investigadoras e Investigadores por M\'exico''. The authors thankfully acknowledge the computer resources, technical expertise and support provided by the Laboratorio Nacional de Superc\'omputo del Sureste de M\'exico, CONACYT member of the network of national laboratories, and by the Laboratorio Nacional de C\'omputo de Alto Desempe\~{n}o (LANCAD), project 13-2021. R.W. \& J.P. acknowledge financial support from the Czech Science Foundation project No. 19-15008S and by the Astronomical Institute of the Czech Academy of Sciences through the project RVO:67985815. D.Sz. has been supported by the Alexander von Humboldt Foundation. 
The authors thank the anonymous Referee for a careful reading and helpful suggestions which greatly improved the paper. 
\\
{\it Software:} FLASH v4.3 \citep{Fryxelletal2000}, Numpy \citet{Numpy}, Wind \citep{Wunschetal2017}, TreeRay \Citep{Wunschetal2018}, \textsc{Cinder} \citep{MartinezGonzalezetal2018}.

\bibliographystyle{apj}
\bibliography{master.bib}

\end{document}